\documentclass[twocolumn]{aa} 

\usepackage{graphicx}
\usepackage{txfonts}

\begin{document}

\def\deg{$^{\rm o}$}

\title{Chandra observations of dying radio sources in galaxy clusters}

\author{M. Murgia \inst{1} \and M. Markevitch\inst{2} \and F. Govoni\inst{1} \and P. Parma\inst{3} 
 \and R. Fanti\inst{3} \and H.R. de Ruiter\inst{3} \and K.-H. Mack\inst{3}}

\institute{
INAF\,-\,Osservatorio Astronomico di Cagliari, Loc. Poggio dei Pini, Strada 54,
I-09012 Capoterra (CA), Italy
\and
NASA Goddard Space Flight Center, Code 662, Greenbelt, MD 20771, USA 
\and
INAF\,-\,Istituto di Radioastronomia, Via Gobetti 101, I-40129 Bologna, Italy
}

\date{Received; Accepted}

\abstract
{The dying radio sources represent a very interesting and largely
unexplored stage of the active galactic nucleus (AGN) evolution. They are considered to be very rare, and
almost all of the few known ones were found in galaxy clusters. However, considering the small 
number detected so far, it has not been possible to draw any firm conclusions about their X-ray environment.}
{We present X-ray observations performed with the Chandra
satellite of the three galaxy clusters Abell 2276,
ZwCl 1829.3+6912, and RX J1852.1+5711, which harbor at
their center a dying radio source with an ultra-steep spectrum that we recently discovered.
}
{We analyzed the physical properties of the X-ray emitting gas surrounding these elusive radio sources. We determined the global X-ray properties
 of the clusters, derived the azimuthally averaged profiles of metal abundance, gas temperature, density,
and pressure. Furthermore, we estimated the total mass profiles.}
{The large-scale X-ray emission is regular and spherical, suggesting
a relaxed state for these systems. Indeed, we found that the
three clusters are also characterized by significant enhancements
in the metal abundance and declining temperature profiles toward
the central region. For all these reasons, we classified RX J1852.1+5711, Abell 2276, and ZwCl 1829.3+6912 as cool-core galaxy clusters.}
{
We calculated the non-thermal pressure of the radio lobes assuming that the radio sources are in the minimum energy condition.
For all the dying sources we found that this is on average about one to two orders of magnitude lower than that of the external
gas, as found for many other radio sources at the center of galaxy groups and clusters. 
We found marginal evidence for the presence of X-ray surface brightness depressions coincident with the fossil radio lobes of the dying sources in 
A2276 and ZwCl\,1829.3+691. We estimated the outburst age and energy output for these two dying sources. The energy power from the AGN outburst is significantly higher than the X-ray luminosity in both clusters. Indeed, it is sufficient that a small fraction of this power is dissipated in the intra-cluster medium to reheat 
the cool cores.
}


\offprints{M. Murgia, matteo@oa-cagliari.inaf.it}

\titlerunning{Chandra observations of dying radio sources in galaxy clusters}
\maketitle

\section{Introduction}
Dying radio sources represent the last stage in the life-cycle of radio
galaxies. During their active stage, which may last several $10^{7}$ years,
the strong synchrotron sources associated with elliptical galaxies are powered 
with energy from the AGN via jets of plasma.
In this stage, the total spectra of the radio sources are usually well approximated by a power law 
over a wide range of frequencies. However, when the activity in the nucleus stops or falls to such a low level that 
the plasma outflow can no longer be sustained, the radio source undergoes a period of fading
(dying phase) before it disappears completely.  In the dying phase, the radio core, well-defined jets, and compact
hot-spots will disappear because they are the structures produced by continuing activity. However,
the radio lobes, which accumulated the relativistic particles produced during the active phase, may still remain 
detectable for a longer time. An example of such a source was given in Cordey (1987). 
It is also possible that radio galaxies may be active intermittently. In this scenario, one expects to observe
fossil radio plasma remaining from an earlier active epoch, along with newly
restarting jets. The best case for fading radio lobes seen with a currently active galaxy
is 3C 338 (see Gentile et al. 2007 for a recent work).

In principle, every radio galaxy must inevitably enter the fading phase. Nevertheless,
dying radio sources are very rare objects. Only a few percent of the 
radio sources in the B2 and 3C samples have the characteristics of a dying radio galaxy (Giovannini et al. 1988). 
A possible explanation for the rarity of dying radio galaxies may be the strong spectral evolution that occurs 
in the source during the fading phase. Without fresh particle injection, the high-frequency
radio spectrum develops an exponential cutoff due to the radiative losses (e.g. Komissarov \& Gubanov 1994).
At this point, the adiabatic expansion of the radio lobes will concur to shift this spectral break to lower frequencies
and the source will quickly disappear from the sky.

However, if the adiabatic expansion of the fading radio lobes 
is somehow reduced, or even stopped after attaining the pressure equilibrium, there is still the chance to detect the
dying source, at least at lowest radio frequencies. Indeed, due to their ultra-steep and curved radio spectra,
dying radio galaxies are more easily detected in low-frequency selected samples. The Westerbork 
Northern Sky Survey (WENSS; Rengelink et al. 1997) at 325 MHz and the B2 survey at 408\,MHz (Colla et al. 1970, 1972, 1973) are particularly
well-suited to search for these elusive fossil radio sources. Parma et al. (2007) discovered nine new dying sources 
by cross-correlating the WENSS with the NRAO VLA Sky Survey (NVSS; Condon et al. 1998). Another two dying galaxies 
(the central radio source in Abell 2622 and MKW03s) and one possibly restarting source (MKW07) were found 
by Giacintucci et al. (2007) in a low-frequency survey of nearby galaxy clusters performed with the Giant Metrewave Radio Telescope.
Five more dying radio galaxies with extremely steep spectra have been found in the WENSS and the B2 catalogs by Murgia et al. (2011).

These studies demonstrate that the dying sources could be among the most promising targets for the upcoming next generation of low-frequency
interferometers. Owing to their high sensitivity and angular resolution, LOFAR, LWA, and in the future SKA, represent the ideal instruments 
to discover and study these elusive objects in detail. 
In the meantime, we can begin to address a few significant points. One is the X-ray environment of the dying sources discovered so far. 
The gaseous environment in which radio galaxies are embedded may play a fundamental role in the later stages of
the radio source life. It seems that there is a tendency for dying sources to reside in dense environments.
Parma et al. (2007) found that about half of the dying sources of their sample are located in galaxy clusters, while only a few appear to be isolated. The five dying galaxies presented in Murgia et al. (2011) are not an exception to the rule: each source is located, at least in projection, at 
the center of an X-ray emitting cluster or galaxy group. The galaxy clusters are Abell 2276, ZwCl\,1829.3+6912, 
RX\,J1852.1+5711, ZwCl\,0107.5+3212, and Abell 2162 for the radio 
sources WNB 1734+6407, WNB 1829+6911, WNB 1851+5707, B2 0120+33, and B2 1610+29. 

\begin{table}[t]
  \begin{center}
  \caption{Coordinates and distances of the three galaxy clusters studied in this work.}

\medskip
  \label{tab:pages}
  \begin{tabular}{ccccc}
\hline
\noalign{\smallskip}
    Cluster & R.A. &  DEC. &  redshift & scale          \\
            & (J2000)& (J2000)    &   &(kpc/\arcsec)  \\ 
\hline
\noalign{\smallskip}
RX\,J1852.1+5711  & 18 52 08 & +57 11 42 & 0.1068  &  1.9 \\
Abell 2276        & 17 35 05 & +64 06 08 & 0.1406  &  2.4 \\
ZwCl\,1829.3+6912 & 18 29 06 & +69 14 06 & 0.204   &  3.3 \\
\hline
\end{tabular}
\end{center}
\label{basic}
\end{table}

Although no firm conclusions can be drawn because of the small number statistics involved,
the results in Murgia et al. (2011) suggest that the probability for a dying radio source to be found
in a cluster is as high as $\sim 86$\%. The simplest interpretation for the 
tendency of dying galaxies to be found in clusters is that the low-frequency radio emission from the fading radio lobes lasts longer 
if their expansion is somewhat reduced or even stopped by the pressure of a particularly dense intra-cluster medium.  
Another possibility is that the occurrence of dying sources is intrinsically higher in galaxy clusters.

To investigate these hypotheses, we need to compare in detail the actual fading radio structures with the properties of the X-ray
emitting gas.  However, the X-ray environment of dying sources is poorly known in general. For instance, only 2 out of a total of 14 dying radio sources in clusters 
presented in Parma et al. (2007) and Murgia et al. (2011) have well-measured X-ray properties. This is not surprising since these samples have been selected originally 
in the radio band and thus adequate X-ray follow-up is required. We had the chance to observe three of these clusters, Abell 2276, ZwCl\,1829.3+6912, and RXC J1852.1+5711, with the Chandra satellite. Very little is known about these galaxy clusters. In this paper we report the results we obtained from the study of the Chandra data.

In Sect.\,2 we described the X-ray observations and data reduction. In Sect.\,3 we present the X-ray analysis results. The implications of these results are 
discussed in Sect.\,4. and a final summary of the conclusions is given in Sect.\,5. In addition, we report the details of the minimum energy calculation  
in the appendix.

Throughout this paper we assume a $\Lambda$CDM cosmology with
$H_0$ = 71 km s$^{-1}$Mpc$^{-1}$, $\Omega_m$ = 0.3, and $\Omega_{\Lambda}$ = 0.7.  Unless otherwise stated, reported uncertainties are at 1$\sigma$ confidence level.

\begin{figure*}
\begin{center}
\includegraphics[width=17cm]{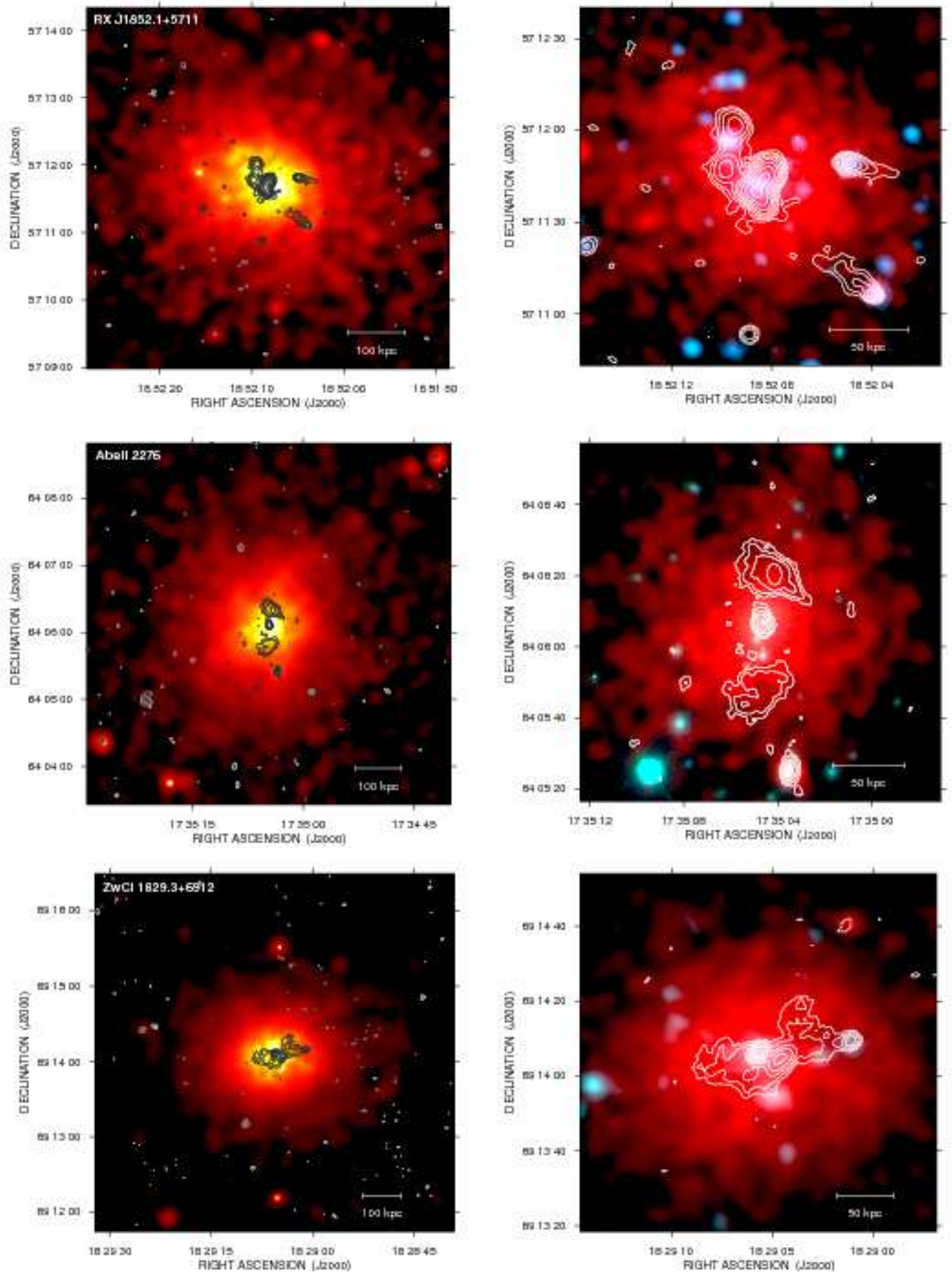}
\caption{Background-subtracted and exposure-corrected Chandra ACIS-I image in the 0.5$-$7 keV 
band of RX\,J1852.1+5711, Abell 2276, and ZwCl\,1829.3+6912. In the left-column panels the adaptively smoothed Chandra images are shown with the VLA radio contours at 1.4 GHz overlaid. A zoom of the central regions of the 
galaxy clusters are shown in the right-column panels. Red tones represent the Chandra image smoothed with a Gaussian kernel with $\sigma=2\arcsec$ while the blue tones represent the optical images from the DSS (RX\,J1852.1+5711 and ZwCl\,1829.3+6912) and SDSS (A2276).}

\end{center}

\label{overlays}
\end{figure*}

\section{Chandra observations and data reduction}

We observed the galaxy clusters Abell 2276, ZwCl\,1829.3+6912, and RX\,J1852.1+5711 with the
Chandra satellite using the ACIS-I detector. The basic properties of the three galaxy clusters are reported in Tab.\,\ref{basic} while 
 the summary of observing exposure times, ID, and dates is given in Tab.\,\ref{obssum}.

The data calibration was performed with the help of the Chandra's data analysis system
 CIAO (Fruscione et al. 2006).
The photon list was filtered for bad events following the standard procedures. In particular, we removed
bad pixels and columns, events with ASCA grades 1, 5, and 7. Data were taken in VFAINT mode and we filtered 
the event file to exclude bad events associated with cosmic ray afterglows. 

We examined the light curve of the X-rays from the portions of the detector free from the cluster or point 
sources. We excluded all time intervals where the background fluctuations were stronger than $\pm 20\%$ of the average.
We discarded a few percent of total exposure time for each cluster.

The background was modeled using a composite blank-field background data set corresponding to the period of
 the observations, cleaned and normalized as described by Markevitch et al. (2003). 

\begin{table}[h]
\caption[]{Chandra ACIS-I observations summary.}
\begin{center}
\begin{tabular}{ccccc}
\hline
\noalign{\smallskip}
   Target&Obs. ID &  Date &  Exposure\\
         &  &       &      ks\\
\noalign{\smallskip}
\hline
\noalign{\smallskip}
RX\,J1852.1+5711    & 5749  & 2006-May-03& 30& \\
Abell 2276         & 10411  & 2009-Jul-10& 40\\
ZwCl\,1829.3+6912   & 10412/10931 & 2009-Jul-22/24& 70\\
\noalign{\smallskip}
\hline
\end{tabular}
\end{center}
\label{obssum}
\end{table}

For each of the three galaxy clusters, we generated an approximate exposure map from the aspect histogram file and the instrument map by assuming
 a mono-energetic distribution of source photons at the same energy as the count rate histogram peak.

\section{X-ray analysis results}
For the data analysis we made use of the XSPEC package (Arnaud 1996).

\begin{table*}[t]
\caption[]{Global spectral properties derived from the fit of the APEC model.}
\begin{center}
\begin{tabular}{cccccccc}
\hline
\noalign{\smallskip}
   Cluster    & Galactic $N_{H}$       &  $N_{H}$            & $kT$       & $Z/Z_{\odot}$  &  $norm$   & $\chi^{2}/d.o.f.$ & $L_{\rm X}$\\
              &   $10^{20}$cm$^{-2}$      &  $10^{20}$cm$^{-2}$ & keV      &                & $10^{-3}$cm$^{-5}$ &  & $10^{44}$ erg\,s$^{-1}$    \\
\noalign{\smallskip}
\hline
\noalign{\smallskip}
RX\,J1852.1+5711 & 4.3   & $4.0\pm1.2$       & $3.30\pm0.15$ & $0.27\pm0.06$ & $2.67\pm 0.11$ & 1.01 (238.7/236) &  $0.72$\\
Abell 2276       & 2.6   & $4.9\pm 1.5$      & $3.14\pm0.17$ & $0.25\pm0.06$ & $1.72\pm 0.08$ & 0.87 (188.6/217) &  $0.77$\\
ZwCl\,1829.3+6912& 6.2   & $10.5\pm 1.4$     & $3.44\pm0.15$ & $0.31\pm0.06$ & $1.58\pm 0.06$ & 0.90 (307.9/346) &  $1.55$\\
\noalign{\smallskip}
\hline
\multicolumn{7}{l}{\scriptsize Col.2: The total Galactic HI column density from the Leiden/Argentine/Bonn (LAB) Survey of Galactic HI (Kalberla et al 2005);}\\
\multicolumn{7}{l}{\scriptsize Col.8: Unabsorbed X-ray luminosity in the 0.5$-$7 keV rest-frame band;}\\

\end{tabular}
\end{center}
\label{globalspec}
\end{table*}

\subsection{Large-scale X-ray emission and global physical properties}
The Chandra X-ray images in the 0.5$-$7 keV band of the three galaxy clusters are shown in Fig.\,1. 
In the left-column panels the adaptively smoothed Chandra images of the large-scale X-ray emission are shown with 
the VLA radio contours at 1.4\,GHz overlaid (see Murgia et al. 2011 for the details about the radio data). The dying radio 
sources are all located at the center of the clusters.

In the right-column panels of Fig.\,1, we show the zoom of the central regions of the galaxy clusters where the blue tones represent the optical images 
from the Digitized Sky Surveys (DSS) and Sloan Digital Sky Survey (SDSS). 

We extracted the global spectra of the three galaxy clusters from a circular region of $R_{E}=250$ kpc in radius centered on the X-ray peak.
We masked the X-ray point sources and fitted the global spectra with a single-temperature APEC (astrophysical plasma emission code; Smith et al. 2001) model 
and Galactic absorption.  We let the total Galactic HI column density ($N_{H}$) vary in the fit procedure. The best-fit parameters are summarized in Tab.\,\ref{globalspec} where, for comparison, we also report the Galactic $N_{\rm H}$ derived by radio surveys\footnote{$N_{\rm H}$ web tool at http://heasarc.gsfc.nasa.gov.}.

\subsubsection{RX\,J1852.1+5711}

The large-scale X-ray emission can be traced out to 
more than about two arcminutes from the cluster center and is quite regular and symmetric in shape. The most striking feature seen in the X-ray image is a strong 
peak of emission at the cluster center. The bright X-ray core hosts the complex radio source WNB1851+5707. Spectral studies based on VLA radio data 
suggest that WNB1851+5707 is in reality composed by two distinct dying radio galaxies: WNB1851+5707a, the amorphous radio source at cluster center, and 
WNB1851+5707b, the fainter double-structure source a few tens of kpc to the N-E of the core (Murgia et al. 2011). The two hosting galaxies have nearly the same redshifts, which additionally supports that they are spatially nearby and not positionally coincident just because of a projection 
effect.
This association is intriguing given the rarity of this kind of sources. 
At the edge of the bright X-ray core two narrow head-tail radio sources, apparently traveling in opposite direction with respect to
 the intra-cluster medium, are visible (see Fig.\,1 top panel).

\subsubsection{Abell 2276}
The overall cluster X-ray emission is very regular and spherical with a clear enhancement at the cluster center. A zoom of the core is presented in the middle-right
 panel of Fig.\,1. Abell 2276 hosts the dying source WNB1734+6407 (Murgia et al. 2011). Characterized by two relaxed lobes lacking jets and hot-spots, 
the radio morphology of WNB1734+6407 can be considered the prototype of fossil radio galaxies.  
The VLA images also reveal a slightly extended core component in coincidence with the cD galaxy at the peak of the X-ray emission. 
This feature has a quite steep radio spectrum and its nature remains unclear. The SDSS image shows a chain of three smaller satellite galaxies trailing the
 cD galaxy about 20 kpc to the south of the core. 

\begin{figure*}
\begin{center}
\includegraphics[width=15cm]{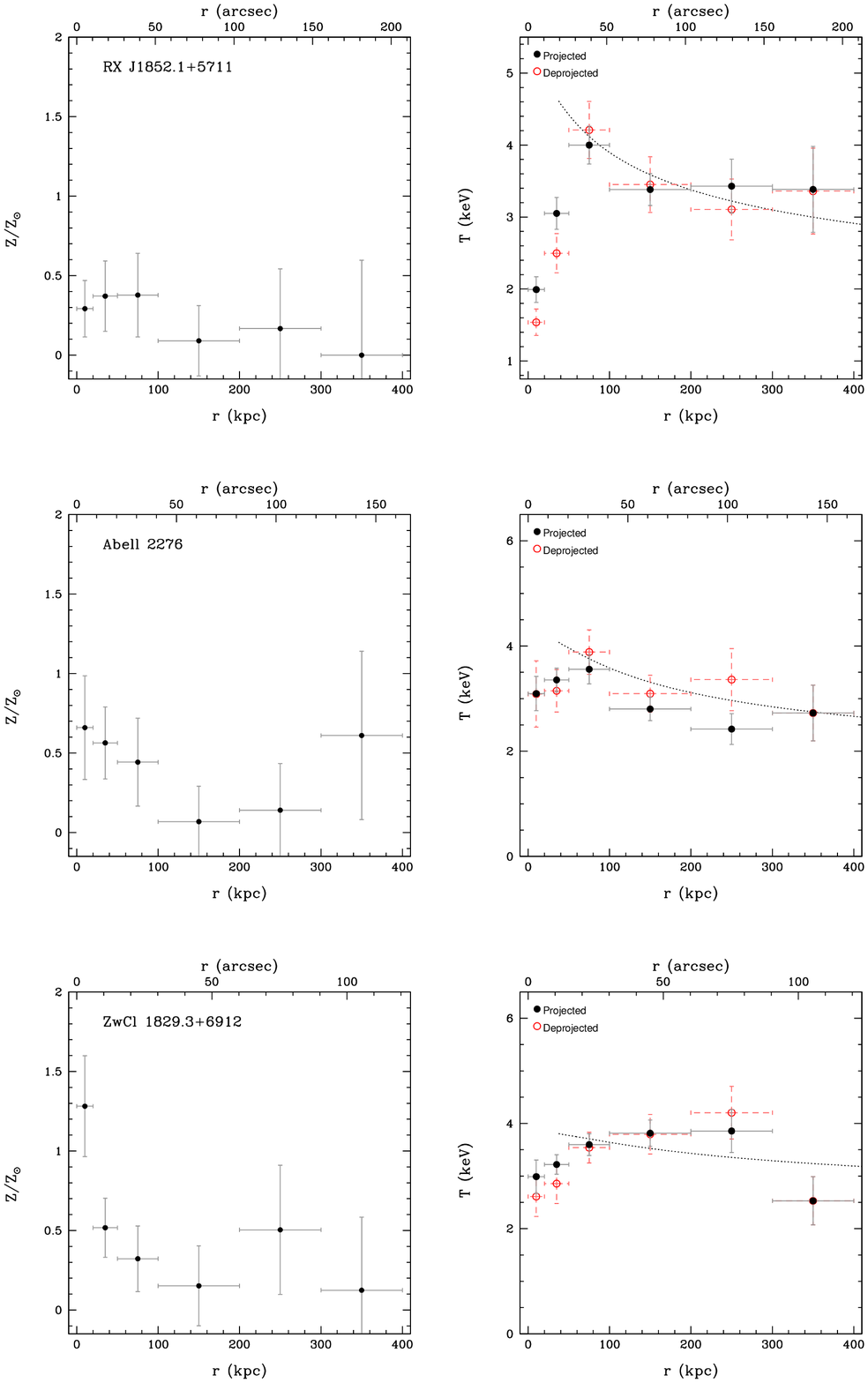}
\caption{ Azimuthally averaged radial profiles of the gas metal abundance (left panel) and temperature (right panel). In the right-column panels the open symbols 
refer to the deprojected temperature profiles while the dotted lines are the best fit of the polytropic model to the four more external annuli, see text.}
\end{center}
\label{fig::radialprofs}
\end{figure*}
\subsubsection{ZwCl\,1829.3+6912}

The large-scale X-ray emission has an ellipsoidal shape, slightly flattened in the N-S direction. A bright peak is present at the center. The cluster hosts the restarting radio source WNB 1829+6911 (Murgia et al. 2011). The radio source is characterized by an active core, associated to the cD galaxy, and a fossil structure extended about 100\,kpc in the E-W direction, following the elongation of the surrounding large-scale X-ray emission.  

\subsection{Metal abundance and temperature profiles}
\label{AnT_prof}
We fitted the metal abundance and the plasma temperature using six concentric annuli centered on the cluster core using an APEC model with the Galactic absorption fixed 
to the value found from fit of the global X-ray spectrum (see Tab.\,\ref{globalspec}). 

The radial profiles of the metal abundance (in the XSPEC default system; Anders \& Grevesse 1989) are shown in the left panel of Fig.\,2,
while the radial profiles of the gas temperature are presented in the right-column panels of of Fig.\,2.

We tried to deproject the temperature profile using the PROJCT model of XSPEC. The statistical quality of our data was not sufficient to permit a detailed 
deprojection of the metal abundance. Therefore, we kept this parameter fixed to the projected value in each annulus. The deprojected temperature profiles are 
represented by the open dots in the right-column panels of Fig.\,2. Although the error bars are large, we note
that toward the cluster center the deprojected temperature profiles are steeper and the gas temperature is lower in the central bins, as expected. 

\begin{figure*}
\begin{center}
\includegraphics[width=18.5cm]{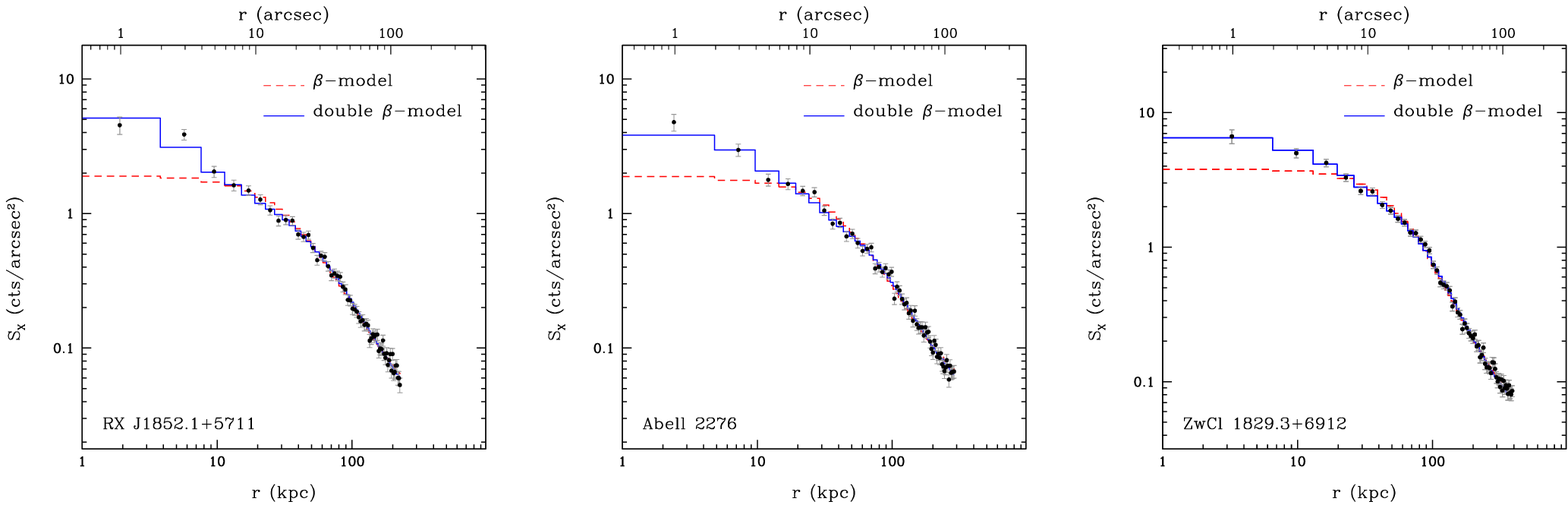}
\caption{Azimuthally averaged $S_X$ profiles in the 0.5$-$7 keV band along with the best fit of the 
isothermal $\beta$-model (dashed line) and double $\beta$-model (continuous line).}
\end{center}
\label{sx}
\end{figure*}

\subsection{Density profile modeling}
\label{densmod}
We extracted the radial profiles of the cluster X-ray surface brightness ($S_X$) by azimuthally averaging the counts in concentric annuli of about $2\arcsec$ in width 
centered on the cD galaxy. We cut out the unrelated sources in the field of view and attempted
to model the observed $S_X$ profiles by first assuming that the intra-cluster medium is in pressure equilibrium and isothermal at the global temperature reported in Tab.\,3.
The temperature radial profiles suggest that the isothermal assumption is applicable for Abell 2276 while it could
be inappropriate to describe the central regions of RX\,J1852.1+5711 and ZwCl\,1829.3+6912. For this reason, at the end of this section we also compare the results of the isothermal model fitting with the deprojection analysis.

Under the assumptions of spherical symmetry, we first attempted to model the observed $S_X$ profiles by assuming the isothermal $\beta$-model for the gas density:

\begin{equation}
n_H(R)=n_{H,0}[1+(R/R_c)^2]^{-3\beta/2},
\label{betamod}
\end{equation}

\noindent
where $R$ is the distance from the cluster center while $R_c$ is the core radius, $\beta$ the index, and $n_{H,0}$ the central density (Cavaliere \& Fusco-Femiano 1976).

The model is calculated in a two-dimensional image with the same pixel size as the observed image. For each pixel in the model image, 
we computed the expected X-ray surface brightness profile by integrating the emission measure along the line of sight:

\begin{equation}
S_X(x,y)\propto  \int_{-\infty}^{+\infty}n_{H}(R)^2 dl,
\label{sxmod}
\end{equation}

\noindent
here $R=\sqrt{r^2+l^2}$, where $r=\sqrt{(x-x_0)^2+(y-y_0)^2}$ is the projected distance from the X-ray peak whose position, $(x_0, y_0)$, has been kept fixed during the fit.

The model is filtered (multiplied) by the mono-energetic exposure map and the background is added before the radial binning is performed. Finally, the expected $S_{X}$ radial profile is compared to the observed one by azimutally averaging the model and the data with the same set of annuli and by minimizing the $\chi^2$ statistic:

\begin{equation}
\chi^2=\sum [(C_{obs}-C_{mod})/Err(C_{obs})]^2,
\label{chi2}
\end{equation}

\noindent
here $C_{obs}$ and $C_{mod}$ are the observed and model counts in each annulus, respectively, and we summed over all annuli.

Since we fixed the position of the X-ray centroid, the $\beta$-model has three remaining free parameters: the core radius, the index $\beta$, and the normalization.
The APEC normalization provided by XSPEC is

\begin{equation}
norm_{\rm APEC}=\frac{10^{-14}}{4\pi D_A^2(1+z)^2}  \int n_en_{H} dV ~~~~ {\rm cm^{-5}},
\label{normapec}
\end{equation}
where $D_A$ is the angular distance to the source. Indeed, we can obtain the central gas
density $n_{H,0}$ from APEC normalization of the global spectrum listed in Tab.\,\ref{globalspec} and the
 values of $\beta$ and $R_c$ found from the fit of the $S_{X}$ profile, by computing the emission measure integral $EI=\int n_en_{H} dV$:

\begin{eqnarray}
&&EI=1.17\cdot 4\pi \int_{0} ^{\pi/2} \int_0^{R_E/\sin\theta} n_H(R)^2R^2dR\sin\theta d\theta=\\ \nonumber
&&=1.17\cdot 4\pi\,n_{H,0}^2\int_{0} ^{\pi/2} \int_0^{R_E/\sin\theta} \left(1+\frac{R^2}{R_c^2}\right)^{-3\beta}R^2dR \sin\theta d\theta,
\label{volumeintegral}
\end{eqnarray}
\noindent
where $R_E=250$ kpc is the radius of the extraction region of the global spectrum and we
assumed that $n_e=1.17 n_H$.

The best fits of the $\beta$-model to the observed $S_X$ profiles are presented as dashed lines in Fig.\,3 while the 
best-fit parameters are listed in Tab.\,\ref{fitpro}.

While it is clear that the $\beta$-model gives a good representation of the 
$S_X$ profiles in the external regions of the clusters, it is far from a good fit in the inner 10 kpc.
The brightness excess of the cool cores of the clusters we considered in this work cannot be described by 
the simple $\beta$-model. Indeed, we also consider the empirical double $\beta$-model presented by Pratt \& Arnaud (2002):

\begin{eqnarray}
R<R_{\rm cut}&&n_H(R)=n_{Hin,0}[1+(R/R_{c,in})^2]^{-3\beta_{in}/2}\\ 
R>R_{\rm cut}&&n_H(R)=n_{Hout,0}[1+(R/R_{c,out})^2]^{-3\beta_{out}/2}.\nonumber
\label{doublebeta}
\end{eqnarray}

The model assumes that both the inner and outer gas
density profile can be described by a $\beta$-model, but with different
parameters. However, the density profile (and its derivate) must be continuous across
 $R_{cut}$ to ensure the continuity of the total mass profile. From this condition
it follows that

\begin{equation}
n_{Hout,0}=n_{Hin,0}\frac{[1+(R_{cut}/R_{c,in})^2]^{-3\beta_{in}/2}}{[1+(R_{cut}/R_{c,out})^2]^{-3\beta_{out}/2}}
\end{equation}

\noindent
and
 
\begin{equation}
\beta_{in}=\beta_{out}\frac{1+(R_{c,in}/R_{cut})^2}{1+(R_{c,out}/R_{cut})^2}.
\end{equation}

We fitted the double $\beta$-model by varing $R_{cut}$, $\beta_{out}$, $R_{c,in}$, and 
$R_{c,out}$, and the model normalization. 
Similarly as for the single $\beta$-model, the value of the central 
gas density, $n_{Hin,0}$, is obtained through the APEC normalization of the global spectrum in Eq.\,\ref{normapec} and the volume integral in Eq.\,5.

The best fits of the double $\beta$-model to the observed $S_X$ profiles are presented as continuous lines in Fig.\,3 while the 
best-fit parameters are listed in Tab.\,\ref{doublefitpro}. The double $\beta$-model gives a good representation of the 
$S_X$ profile also in the inner central 10 kpc and represents a statistically significant improvement over the single $\beta$-model.

The internal core radius in RXJ1852.1+5711 is of about $R_{c,in}\simeq 2$ kpc, while it is somewhat larger in A2276 and ZwCl 1829.3+6912, $R_{c,in}\simeq 8$ and 9 kpc, respectively. 
The rapid increase in the gas density observed for $R<R_{c,in}$ could be interpreted as the result of the cooling flow process where the dense intra-cluster medium
may have already mixed with the progenitor X-ray coronas associated to the cD galaxies (Sun et al. 2007).

\begin{table*}[t]
\caption[]{Results of the surface brightness profile fit of the $\beta$-model.}
\begin{center}
\begin{tabular}{ccccc}
\hline
\noalign{\smallskip}
   Cluster          &  $n_{H,0}$  &   $R_{c}$ &   $\beta$& $\chi^{2}/d.o.f.$ \\
                    &  cm$^{-3}$    &    kpc   &         &                \\
\noalign{\smallskip}
\hline
\noalign{\smallskip}
 RX\,J1852.1+5711   &        $0.015\pm 0.001 $         &  $29.8\pm 0.5$ & $0.48\pm 0.01 $  & 1.80 (102.4/57)  \\
 A2276             &         $0.012\pm 0.001$       &  $39.6\pm 0.8$ & $0.50\pm 0.01$  &  1.82 (103.9/57)  \\
 ZwCl\,1829.3+6912  &        $0.0129\pm 0.0008 $       &  $59.9\pm 0.8$ & $0.60\pm 0.01 $  & 1.59 (89.5/57)  \\
\noalign{\smallskip}
\hline
\end{tabular}
\end{center}
\label{fitpro}
\end{table*}

\begin{table*}[t]
\caption[]{Results of the surface brightness profile fit of the double $\beta$-model.}
\begin{center}
\begin{tabular}{ccccccc}
\hline
\noalign{\smallskip}
   Cluster          &  $n_{H,0in}$  &   $R_{c,in}$ &  $R_{cut}$&  $R_{c,out}$&  $\beta_{out}$& $\chi^{2}/d.o.f.$ \\
                    &  cm$^{-3}$    &    kpc      &   kpc      &    kpc      &               & \\
\noalign{\smallskip}
\hline
\noalign{\smallskip}
 RX\,J1852.1+5711   &        $ 0.026\pm 0.003$       &  $2.1\pm 0.1$ & $24.8\pm 0.3$ & $30.0\pm 0.2$ & $0.41\pm 0.01$  & 0.88 (48.2/55)  \\
 A2276             &         $ 0.015\pm 0.001$       &  $8.0\pm 0.3$ & $44.2\pm 0.6$ & $55.2\pm 0.5 $& $0.50\pm 0.01$  & 0.99 (54.8/55)  \\
 ZwCl\,1829.3+6912  &        $ 0.017\pm 0.001$       &  $9.4\pm 0.3$ & $46.8\pm 0.7 $ &$78.7\pm 0.6$ & $0.64\pm 0.01$  & 0.87 (47.8/55)  \\
\noalign{\smallskip}
\hline
\end{tabular}
\end{center}
\label{doublefitpro}
\end{table*}

These calculations are based on the assumption that the
emissivity of the gas simply scales as density squared. To estimate the effect of temperature variations on the
value of the electron density, we calculated the electron number density
directly from the deprojected spectral analysis (Sec.\,\ref{AnT_prof}) and compared
this with the model electron number density obtained from the fit of the 
isothermal models to the X-ray surface brightness. The left-column panels of Fig.\,4 show that 
the deprojected electron density (represented by the open dots) agrees remarkably well with that obtained from the model analysis.

The gas pressure can be related to density and temperature via the ideal gas law:

\begin{equation}
P\simeq (2n_H+3n_{H_e}) kT = 2.25 n_H kT,
\label{gaspress}
\end{equation}

where we neglected the contribution from elements heavier than helium and adopted $n_{H_e}/n_H=0.083$ for the number density of helium relative to
hydrogen, which is obtained from the primordial mass fraction of helium and by assuming the abundance of heavier elements to be 0.3$-$0.5 solar.

In the right-column panels of Fig.\,4, we compare the deprojected gas pressure  
with the prediction of the isothermal models. The gas pressure is derived for both the $\beta$-model and the double $\beta$-model by assuming for the gas temperature the global values reported in Tab.\,\ref{globalspec}. 
\begin{figure*}
\begin{center}
\includegraphics[width=15cm]{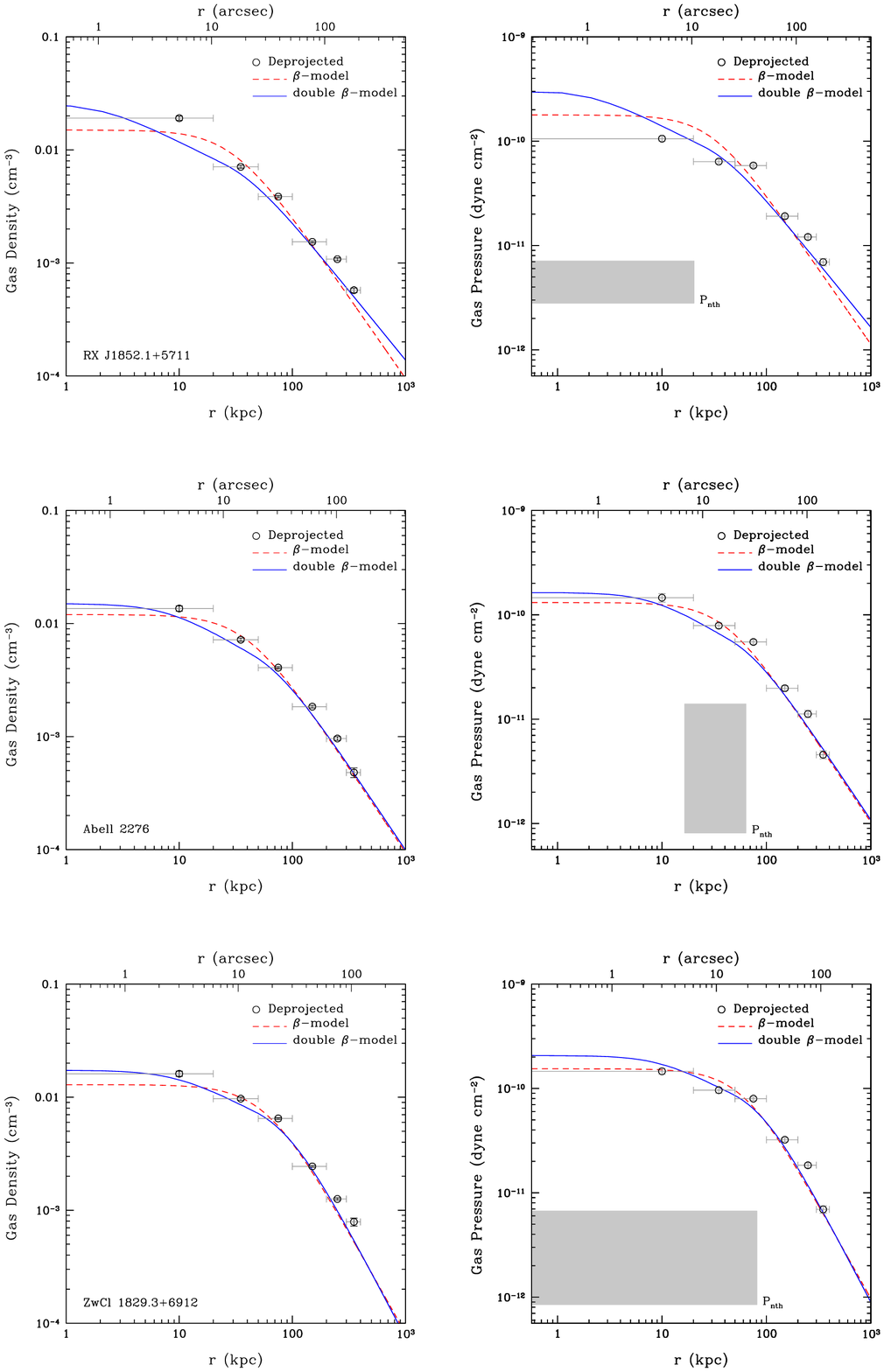}
\caption{Radial profiles of gas density (left-column panels) and pressure (right-column panels). The open dots represent the result of the deprojected spectral 
analysis while the  dashed and continuous line represent the trend of the isothermal $\beta$-model and double $\beta$-model, respectively. The shaded areas represent the 
 range of non-thermal pressure of the radio lobes under the minimum energy assumption, see Sect.\,4.2.}
\end{center}
\label{densnpress}
\end{figure*}

We note that the gas pressure at the cluster center predicted by isothermal models is higher than the deprojected value in RX\,J1852.1+5711 and  ZwCl\,1829.3+6912.
In fact, for these two clusters we observe a much stronger drop of the temperature in the central bins. For Abell 2276 the temperature drop at the center is
 less prominent and thus we observe for this cluster a better agreement between the pressure profiles derived from the isothermal model fit and the values obtained through the deprojection analysis. 
\subsection{Total mass profiles}
We calculated the profiles of total mass under the assumption of hydrostatic equilibrium. For a spherically symmetric distribution of gas the total mass profile
 is given by
\begin{equation}
M(r)=-\frac{kT r}{\mu m_{p} G} \left( \frac{d \ln n_{H}}{d \ln r}-\frac{d \ln T}{d \ln r} \right),
\label{totalmass}
\end{equation}
where $m_{p}$ is the proton mass, $G$ the gravitational constant, and $\mu=0.6$. If the gas density profile is described by the $\beta$-model

\begin{equation}
M(r)=-\frac{k r^2}{\mu m_{p} G } \left( -\frac{3 \beta r T(r)}{r^2+r_{c}^2}+\frac{d T}{d r} \right),
\label{totalmassbeta}
\end{equation}

while for the double $\beta$-model we have

\begin{eqnarray}
M(r<R_{cut})&=&-\frac{k r^2}{\mu m_{p} G } \left( -\frac{3 \beta_{in} r T(r)}{r^2+r_{c, in}^2}+\frac{d T}{d r} \right)\\ 
M(r\ge R_{cut})&=&-\frac{k r^2}{\mu m_{p} G } \left( -\frac{3 \beta_{out} r T(r)}{r^2+r_{c, out}^2}+\frac{d T}{d r} \right). \nonumber
\label{totalmassdoublebeta}
\end{eqnarray}

The total mass profiles for the isothermal models are shown in Fig.\,5. We also calculated the mass profile from the deprojected gas temperature
 and density obtained in Sect. 3.3. We excluded the cool core from the total mass calculation and 
evaluated the radial temperature gradient, $dT/dr$, by fitting a polytropic relation of the form 

\begin{equation}
T(r)=T_{0} \left ( \frac{n_{H}(r)}{n_{H,0}} \right )^{\rho-1}
\end{equation}
to the deprojected temperature measured in the four external annuli, where for the density profile $n_{H}(r)$ we adopted the best fit of the double $\beta$-model. 
The fit is shown as a dotted curve in the right-column plots of Fig.\,2. We found a polytropic index of $\rho=$1.17, 1.15, and 1.06 for 
 RX\,J1852.1+5711, Abell 2276, and ZwCl\,1829.3+6912. The total masses derived from the deprojected temperature profile are shown as dots in Fig.\,5.
The profiles of total mass derived from the two isothermal models are very similar and they both agree well with the masses estimated from the polytropic temperature 
 model in the four external annuli. Extrapolating the profiles up to $r_{500}$ (i.e. the radius at which the mean mass density is 500 times the 
critical density), we found a total mass of 1.0, 1.3, and 2.0 $\times 10^{14} M_{\odot}$ for 
 RX\,J1852.1+5711, Abell 2276, and ZwCl\,1829.3+6912, respectively. These values are in roughly agree with the total mass observed for
 galaxy clusters of this temperature (e.g. Finoguenov et al. 2001).

\section{Discussion}

\begin{figure*}[t]
\begin{center}
\includegraphics[width=18cm]{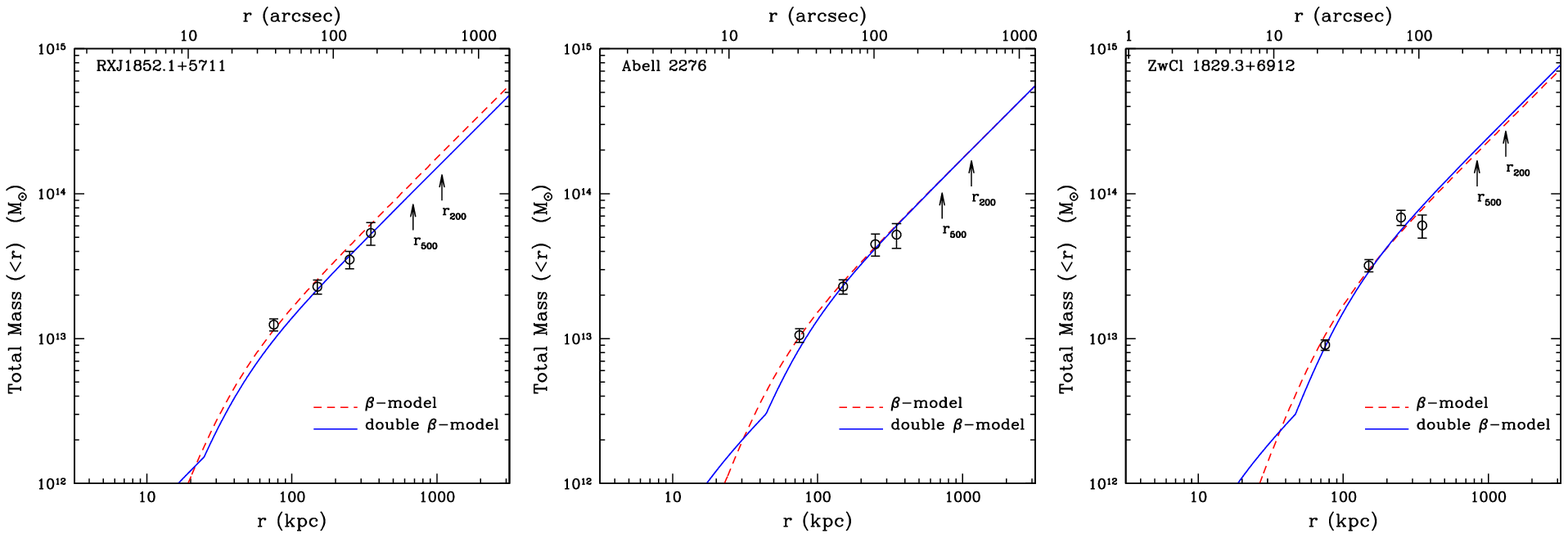}
\caption{Total mass profiles for the isothermal $\beta$-model and double $\beta$-model shown as dashed and continuous lines, respectively. The dots represent the total mass profile derived from the measured deprojected temperature profile by fixing the density profile to the best fit of the double $\beta$-model. The arrows indicate the radii at overdensities of 500 and 200 times the critical density.}
\end{center}
\label{masses}
\end{figure*}
We characterized with data from the Chandra satellite the X-ray emission of three clusters from the Murgia et al. (2011) sample: 
RX J1852.1+5711, Abell 2276, and ZwCl 1829.3+6912, which host a their center the dying sources WNB 1851+5707, WNB 1734+6407, and WNB 1829+6911. 
From the X-ray analysis presented in Sect.\,3, we can conclude that the physical state of the intra-custer medium in the three galaxy clusters presents quite similar properties.
The large-scale X-ray emission is regular and spherical, suggesting a relaxed state for these systems. The most striking feature revealed from the Chandra images 
is a peak of emission at the center of the clusters that cannot be parametrized by the isothermal $\beta$-model. In fact, we found that the three systems
are also characterized by strong enhancements in the central abundance and declining temperature profiles toward the central region. For all these reasons,
 we classify RX J1852.1+5711, Abell 2276, and ZwCl 1829.3+6912 as cool-core galaxy clusters. 
We used the density profile derived in Sec.\,\ref{densmod} together with the deprojected gas temperature to estimate the central cooling time as

\begin{equation}
t_{cool}=2.9\times 10^{10} \sqrt{\frac{kT}{\rm 1\,keV}} \left(\frac{n_{H}}{10^{-3}{\rm cm^{-3}}}\right)^{-1}~~~(\rm yr)
\label{cooltime}
\end{equation}

\noindent
from Pratt \& Arnaud (2002).
For RX J1852.1+5711, Abell 2276, and ZwCl 1829.3+6912 we found a central cooling time of $t_{cool}=$1.3, 3.3, and 2.7 Gyr.
According to
Hudson et al. (2010), the central cooling time is the best parameter to segregate cool-core from non-cool-core clusters. In particular, these authors found from their statistical
 analysis results that clusters whose central cooling time is shorter than the critical value $t_{cool}<7.7$ Gyr are classifiable as cool-cores. 
Indeed, given their comparatively short central cooling time, we can confirm that all three clusters studied here belong to the category of 
the cool-core systems.

It is of interest to compare these results with the X-ray properties of galaxy clusters that host dying sources known in the literature. The prototypical
restarting radio source 3C 338 is associated to the cD galaxy of nearby rich cluster Abell 2199 ($z=0.0309$). The cluster has been studied with Chandra by
 Johnstone et al. (2002), who found evidence for radial gradients in temperature and metallicity in the X-ray emitting gas. The temperature decreases toward
 the central region of the cluster, which is where the radiative cooling time drops below 1 Gyr. As seen in Chandra images, the radio lobes
associated with the AGN significantly affect the X-ray gas, inflating cavities or bubbles in the dense intra-cluster medium. Overall, the X-ray properties of 
Abell 2199 are similar to those we found for the more distant galaxy cluster ZwCl 1829.3+6912. The radio morphologies of the two radio sources are also similar: they both posses a restarting nucleus surrounded
 by dying radio lobes produced in an earlier phase of activity. Among the dying sources studied by Murgia et al. (2011) there is B2 0120+33, which is identified with the galaxy NGC 507, the dominant, massive elliptical galaxy of a nearby ($z=0.01646$) group/poor cluster (the so-called Pisces cluster). The X-ray halo of NGC 507 has been studied with ROSAT HRI and Chandra ACIS-S data by Paolillo et al. (2003), who found that the halo core has a complex morphology with a main X-ray emission peak, coincident with the center of the optical galaxy, and several secondary peaks. The authors found that the energy input by the central 
radio source was strong enough to prevent gas cooling. 
Indeed, the X-ray properties of the clusters examined in this works do not seem to be peculiar compared to those of the few examples already known in the literature. Although the poor statistics prevents a quantitative analysis, there is a tendency for dying sources to be preferentially found in relaxed galaxy clusters. 

\subsection{Are there X-ray cavities associated to the fossil radio lobes?}
It is well established that the central gas in many cool-core systems that host active radio sources is not smoothly distributed, see e.g. 
 the notable cases of Hydra A (McNamara et al. 2000) and MS\,0735+7421 (McNamara et al 2005). The comparison of radio and X-ray images obtained 
at similar angular resolution has revealed that AGN jets are the cause of highly disturbed structures in the cores of many clusters,
 including shocks, ripples, and density discontinuities. The most remarkable structures are the so-called cavities (or radio bubbles), depressions in 
the X-ray surface brightness that are approximately coincident with the radio lobes. The cavities form as the jets propagate in opposite directions outward from the cD galaxy, 
 inflating the lobes of radio plasma and pushing aside the X-ray emitting gas. It is hypothesized that the depressions are nearly devoided of X-ray emitting gas. 
X-ray cavities are common in cool-core clusters; they are present in more than 70\% of these systems (Dunn et al. 2005). There are, hoverer, ``ghost cavities'', i.e., X-ray depressions 
with no detectable radio emission. The interpretation is that these structures have been created by the AGN in the more distant past and their radio emission had faded over the
time (B\^irzan et al. 2004).
In the view of these findings, it is important to discuss if there is evidence of X-ray cavities associated to our dying radio sources. The fading lobes are not powered by the AGN anymore, but their radio emission is still barely detectable. Indeed, dying sources might be an intermediate evolutionary stage in between active and 
ghost cavities.

The galaxy clusters in our study are somewhat more distant compared to the clusters in which most X-ray cavities have been found so far. Thus, detecting the X-ray 
depression is difficult because of the low contrast of our images. 
For this reason, we have produced unsharp mask images by subtracting from the observed counts images the best fit of the double-$\beta$ model to the radially averaged 
$S_{X}$ profiles found in Sect.\,3.3. Fig.\,6 shows the results 
obtained from the 0.5$-$7 keV band by subtracting the circularly symmetric double-$\beta$ model from an image smoothed with a 1\arcsec~Gaussian kernel  then dividing by the sum of the two. Point sources were removed from the images and replaced with values interpolated from surrounding background regions using the CIAO task dmfilth. 
There is no evidence for cavities associated to the radio lobes of WNB 1851+5707 in the unsharp mask image of RX J1852.1+5711 shown in the left panel of Fig.\,6.
The unsharp image of Abell 2276 is presented in the middle panel of Fig.\,6. For this cluster there is a hint of an X-ray depression associated to
 the south lobe of WNB 1734+6407. Finally, the  unsharp image of ZwCl 1829.3+6912 is shown in the right panel 
of Fig.\,6. There is marginal evidence for a cavity in correspondence to the west lobe of WNB 1829+6911. The depression seems roughly to follow the morphology of the radio lobe.
We considered whether we can detect a cavity given the sensitivity of our observations. For that purpose, 
we simulated the expected X-ray depressions that correspond to the radio lobes of WNB 1734+6407 in A2276. We chose this cluster for the simulations 
because the geometry of the problem is simpler to study. We assumed that the gas distribution in the cluster atmosphere is
described by the double-$\beta$ model found in Sect.\,3. Furthermore, we assumed that the fossil lobes of the dying source have an ellipsoidal
 shape and are completely devoid of X-ray emitting material. The results are presented in Fig.\,7 for three inclinations of the 
radio source with respect to the observer. The top panels shows the expected depressions in the X-ray surface brightness for $i=0\degr$ (left-column panels),
  $i=22.5\degr$ (middle-column panels), and $i=45\degr$ (right-column panels). If the source lies on the plane of the sky ($i=0\degr$), the cavities are intrinsically
closer to the cluster's core and thus they are more prominent since they remove dense gas. If the radio source is inclined toward the observer, the cavities are at 
 a larger intrinsic distance from the core, the removed gas is more rarefied and thus the cavities appear less prominent. We quantified the expected fractional depression
 in the X-ray counts over the area of the radio lobes. We constructed synthetic count images by sampling from a Poisson distribution whose mean is given 
by the models shown in the top panel of Fig.\,7. We filtered the synthetic images by the exposure map, added the blank-field background counts, and calculated an 
 expected fractional depression in the X-ray counts of 22\%, 15\%, and 9\% for $i=0\degr$, $22.5\degr$, and $45\degr$. We compared these numbers with 
the observed counts over the region of the south lobe of WNB 1734+6407. For the north lobe the situation is less clear, but this  region of the ACIS-I chip is heavily affected by 
bad columns. Over the region of the south lobe, we measured a fractional depression in the X-ray counts of $19\pm 13$\%, which is 
compatible with the predicted value for the source aligned to the plane of the sky. We note, however, that the uncertainty in the source counts is large and 
thus it is difficult to draw a definitive conclusion. For a less favorable orientation of the radio source ($i>20\degr$) the cavities would be hardly discernible in our image. This is also illustrated in the bottom panels of Fig.\,7, where we show the unsharp mask images corresponding to the three simulated values for the source's inclination. 

We did not perform any simulations for ZwCl 1829.3+691. However, for the west lobe of the dying source WNB 1829+6911 we measured a depression in the X-ray 
counts of $13\pm 7$\%. The X-ray cavity appears to follow the shape of the radio lobe. For the east lobe no statistical significant depression is detected.

In summary, we investigated the presence of X-ray cavities associated to the radio lobes of our dying sources. We found hints of possible 
cavities in A2276 and ZwCl 1829.3+691, but the evidence is weak. However, while cavities are not clearly seen, their presence cannot be excluded either. For the case of A2276, we showed through simulations that if the radio source is on the plane of the sky, the effect of the cavities is consistent with the faint depressions observed in the X-ray surface brightness while the cavities would be hardly discernible in our image if the radio source were significantly inclined toward the observer. Additional, more sensitive observations are necessary to confirm the presence of these features unambiguously.

\begin{figure*}[t]
\begin{center}
\includegraphics[width=18cm]{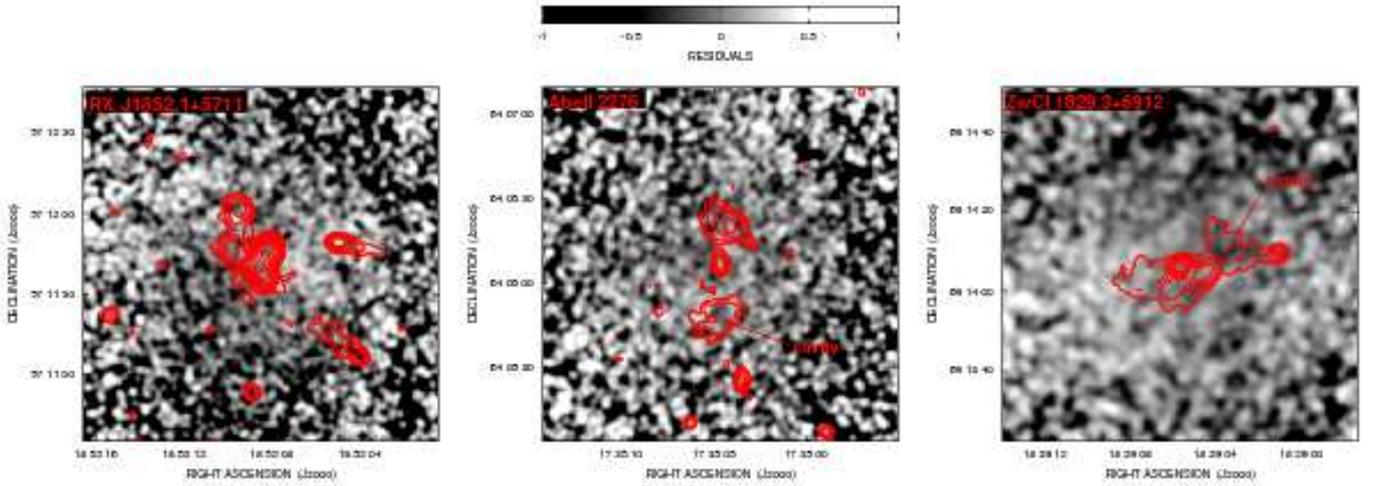}
\caption{Unsharp mask images obtained from the 0.5$-$7 keV band by subtracting the best fit of the circularly symmetric double-$\beta$ model
from the count image smoothed with a 1\arcsec\,Gaussian kernel, and then dividing by the sum of the two.}
\end{center}
\label{unsharp}
\end{figure*}

\begin{figure*}
\begin{center}
\includegraphics[width=18cm]{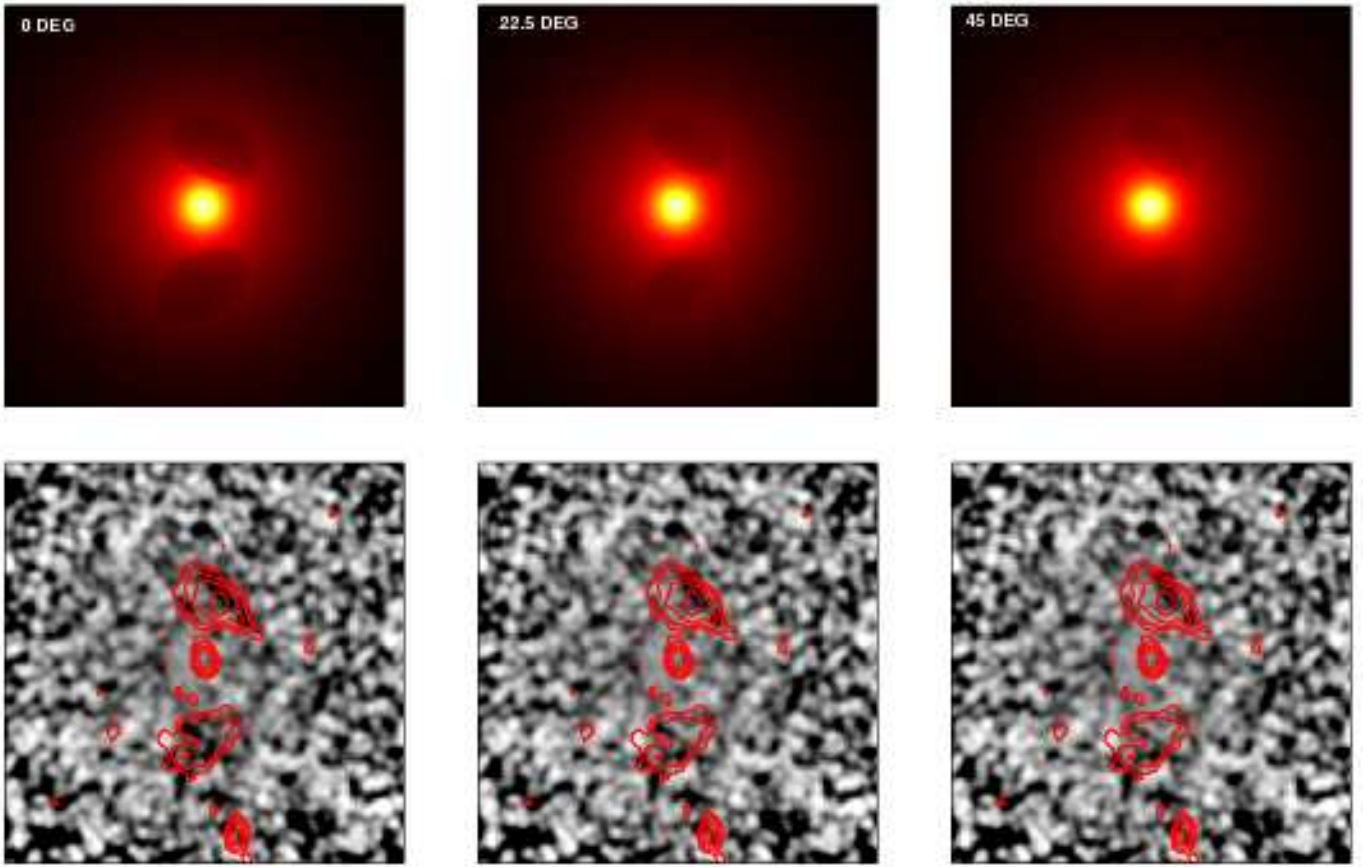}
\caption{Simulated cavities in A2276 for the double-$\beta$ model (top panels). The synthetic unsharp mask images are shown in the bottom panels. We expect
 a depression in the X-ray surface brightness of 22\%, 15\%, and 9\% for $i=0\degr$, $22.5\degr$, and $45\degr$.}
\end{center}
\label{simul}
\end{figure*}

\subsection{Pressure balance}
The gas pressure in the immediate regions of the dying sources is close to $P_{th}\sim 10^{-10}$ dyne/cm$^{-2}$ for all three clusters.

We calculated the non-thermal pressure of the radio lobes as

\begin{equation}
P_{nth}=(\Gamma_c-1)(u_{ions}+u_{el}+u_{B}),
\end{equation}

\noindent
where $\Gamma_c=4/3$ is the adiabatic index of the material in the lobes while $u_{ions}$, $u_{el}$, and $u_{B}$ are the energy densities of relativistic ions, electrons, and
 magnetic field. Under the hypothesis that the radio source is in a  minimum energy condition,  we can compare the non-thermal pressure
of the radio plasma, $P_{nth}=u_{min}/3$, to that of the external medium. The electron energy spectrum is taken to be a power law whose slope is derived from the spectral modeling presented in Murgia et al. (2011) while the volume of the emitting regions is estimated from the radio images assuming a unity filling factor. We also assumed an isotropic electron population spiralling in a completely disordered magnetic field and averaged the synchrotron emissivity over all possible orientations of the field with respect to the line-of-sight (see the appendix for details).
For steep electron energy spectra the electron energy density 
 depends critically on the low-energy cut off, $\gamma_{min}$, whose value is rather uncertain. Traditionally,
 the energy spectrum is truncated at the energy of the electrons radiating at the lower end of the radio band, $\nu_{low}=10$\,MHz (see e.g., Pacholczyk 1970). This choice minimizes the source energetics
required by the observed radiation in the radio band. However, it has been pointed out that it would miss the contribution from lower energy electrons 
(Brunetti et al. 1997, Beck \& Krause 2005). Indeed, we also calculated $u_{el}$ by extrapolating the power law energy spectrum to $\gamma_{min}=1$. Another poorly
 known variable is the relative energy density of relativistic ions to that of electrons, $k=u_{ions}/u_{el}$. We adopted $k=1$ as reference value.

The internal pressure ranges are reported in Tab.\,\ref{minp} and plotted as shaded regions in the right-column panels of Fig.\,4. The lower and upper bound of the range correspond to $\nu_{low}=10$\,MHz and $\gamma_{min}=1$, respectively. For all the dying sources, we found that $P_{nth}$ is on average about one to two orders of magnitude lower than that of the external gas. The gap is reduced for $\gamma_{min}=1$, but still remains significant, especially
 for the dying source in RX\,J1852.1+571 and ZwCl 1829.3+691.

This discrepancy is known for a long time (e.g. Morganti et al. 1988) and has been found for many other radio 
sources at the center of galaxy groups and clusters (Croston et al. 2008). Several hypotheses can be put forward to explain why $P_{th}>P_{nth}$:

\begin{itemize}
\item[(i)] the source is not in the minimum energy condition;
\item[(ii)] there is a large number of ``uncounted'' electrons radiating at frequencies well below the explored radio band;
\item[(iii)] there is a large contribution to the non-thermal energy by protons or other heavy relativistic particles, i.e., $k\gg 1$;
\item[(iv)] ambient thermal material entered the radio lobes, providing an additional source of pressure;
\item[(v)] the volume filling-factor of the radio emitting plasma is smaller than unity.
\end{itemize}

Distinguishing among these possibilities is not trivial since we cannot even exclude that they are acting in combination. 
However, one can attempt to discuss them separately. Scenario (i) require that either the magnetic field energy density or the relativistic particle 
energy density are much higher than the equipartition value. To sustain an external pressure as high as $P_{th}\sim 10^{-10}$ dyne/cm$^{-2}$, a magnetic field strength of 90 $\mu$G is needed. This would imply extremely severe radiative losses for the synchrotron electrons. Especially for a dying source, where the energy supply from the AGN is switched off, the electron energy spectrum will be burned out rapidly in the absence of any compensating re-acceleration mechanism. Indeed, this scenario could be
potentially less problematic if the deviation from equipartition is caused by a higher energy density in relativistic particles.

Scenario (ii) postulates the presence of a bulk of low-energy electrons in the source that radiate below the observable radio band.
We already showed that extending the energy spectrum down to $\gamma_{min}=1$ is not sufficient to fill the pressure gap.
Thus, for this scenario to work, there would have to be a pile-up of low-energy electrons further in excess of the extrapolation of the power law energy spectrum. On the other hand,
 according to scenario (iii) the radio lobes may contain a high percentage of relativistic protons. To fill the pressure gas it is required that $k\sim 100\div 1000$. 
These values are in the range found for central sources in clusters (B\^irzan et al. 2008) and suggest that these radio lobes have been inflated by heavy jets.
However, the matter content of AGN jets is still extremely controversial and the various attempts to determine the jet composition by indirect means led to contradictory conclusions (see e.g. Marsher et al. 2007). According to scenario (iv) there could be a significant amount of thermal material inside the radio lobes to account for a large
 part of the total pressure. The entrainment of surrounding material may occur in the radio jets during the active phase (Laing et al. 1999), and 
it cannot be ruled out that sources close to pressure balance, such as our dying sources, could be more susceptible to turbulence and higher entrainment rates.
Finally, as outlined in scenario (v), the radio emitting volume could be much lower than assumed either because the magnetic fields is highly filamentary and/or the 
shape of the radio lobes is not ellipsoidal but rather flattened to a lenticular shape or bent in a toroidal shape as they rise through the external medium (Churazov et al. 2001).

\begin{table*}[t]
\caption[]{Non-thermal pressure of radio lobes in the three galaxy clusters assuming minimum energy.}
\begin{center}
\begin{tabular}{cccccccc}
\hline
\noalign{\smallskip}
                          &                &         &               &  \multicolumn{2}{c}{$\nu_{low}=10$ MHz}              & \multicolumn{2}{c}{$\gamma_{min}=1$}\\
    Cluster               &$L_{150}$(W/Hz) & $\alpha$ & $V$ (kpc$^3$)&  $B_{min}$ ($\mu$G)     &   $P_{nth}$ (dyne/cm$^2$)  &  $B_{min}$ ($\mu$G)      &    $P_{nth}$ (dyne/cm$^2$) \\
\noalign{\smallskip}
\hline
\noalign{\smallskip}
RX\,J1852.1+571        &1.8$\times 10^{25}$ & 0.5 & 8.4$\times 10^3$ & 9.6  &  2.8$\times 10^{-12}$  & 15.1   &  7.1$\times 10^{-12}$\\
A2276                  &2.6$\times 10^{25}$ & 0.9 & 2.4$\times 10^5$ & 5.5  &  8.1$\times 10^{-13}$  & 22.7   &  1.4$\times 10^{-11}$\\      
ZwCl\,1829.3+6912      &7.6$\times 10^{25}$ & 0.7 & 2.7$\times 10^5$ & 5.4  &  8.5$\times 10^{-13}$  & 15.2   &  6.7$\times 10^{-12}$\\
\noalign{\smallskip}
\hline
\multicolumn{8}{l}{\scriptsize Cols. 1 and 2: Radio source luminosity at 150 MHz and spectral index, taken from Murgia et al. (2011);}\\
\multicolumn{8}{l}{\scriptsize Cols. 5 and 6: Minimum energy magnetic field and non-thermal pressure assuming $\nu_{low}=10$ MHz and $k=1$;}\\
\multicolumn{8}{l}{\scriptsize Cols. 7 and 8: Minimum energy magnetic field and non-thermal pressure assuming $\gamma_{min}=1$ and $k=1$.}\\
\end{tabular}
\end{center}
\label{minp}
\end{table*}

\subsection{Outburst age and energetics}
The injection of energy by the AGN is the favored solution to the cooling problem in relaxed galaxy clusters.
The synchrotron plasma is less dense than the intra-cluster medium and so the radio lobes are expected to detach from the
core and rise up buoyantly through the cluster (e.g. Dunn \& Fabian 2006). It is well known that 
the energy radiated away through the synchrotron emission is only a tiny fraction of the total outburst energy. 
The mechanical work done by the radio source dominates the outburst energetic and can be inferred directly for those 
clusters where an X-ray cavity is detected in correspondence of the radio lobe. Indeed, we attempted to estimate the
 outburst energetics of the dying sources  in Abell 2276 and ZwCl 1829.3+691, the two clusters for which we have
 a putative cavity detection. If radio lobes are inflated sub-sonically, the energy necessary to form the cavity can be evaluated as 
the sum of the internal energy and the work done against $P_{th}$, the external pressure of the thermal gas,

\begin{figure}[t]
\begin{center}
\includegraphics[width=9cm]{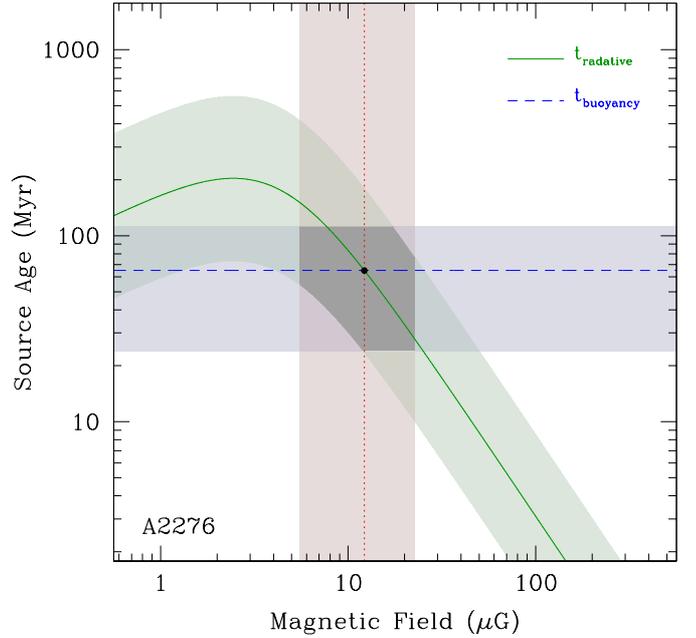}
\caption{Comparison of radiative and buoyancy ages as a function of the magnetic field 
for the dying source WNB\,1734+6407 in A2276. The radiative age (green line and shaded region) is derived from the spectral break measured by Murgia
 et al. (2011). The horizontal blue dashed line and shaded region is the buoyancy age for the south lobe. The two ages match at 65 Myr for $B\simeq 12 ~\mu$G. 
This value for the source's magnetic field (dotted line) fits within the equipartition range for $k=1$ pictured by the vertical shaded region in red.}

\end{center}
\label{tradvstb}
\end{figure}

\begin{equation}
E=\frac{\Gamma_c}{\Gamma_c-1} P_{th}V,
\end{equation}

\noindent
where $V$ is the volume of the cavity while $\Gamma_c$ is the adiabatic index of the fluid in the radio lobe. For a relativistic
fluid $\Gamma_c=4/3$ the outburst energy contained in the cavity is $E=4 P_{th}V$.  To estimate the rate at which the AGN energy
input is dissipated into the intra-cluster medium we need to estimate the age of the outburst.

In A2276 the dying lobes are detached from the AGN and are likely rising in the cluster atmosphere pushed by the buoyant force. 
Assuming that the density of the relativistic fluid in the lobe is negligible with respect to that of the external medium, 
the buoyancy velocity of the lobes can be estimated as 
 
\begin{equation}
v_{b}\sim \sqrt{g\frac{V}{S}\frac{2}{C_{D}}},
\end{equation}
where $g$ is the gravitational acceleration, $S$ is lobe cross section, and $C_{D}$ the drag coefficient (Churazov et al. 2001).

For the putative cavity that corresponds to the south radio lobe in A2276 we assumed an ellipsoidal geometry with major and minor semi-axes of $a=30$ and $b=21$ kpc, respectively. 
We calculated the gravitational acceleration from the total mass profile of the cluster  (see Sect.\,3.4) and, from the volume $V=(4/3) \pi a b^2$ and the cross-section $S=\pi a b$ of the cavity, we found $v_{b}\simeq 600$ -- 850 km/s for $C_{D}\approx$ 1.0 -- 0.5.
 
The sound speed in the intra-cluster medium is

\begin{equation}
c_{s}= \sqrt{\Gamma_{gas} \frac{kT}{\mu m_{H}}},
\end{equation}

\noindent
where $\Gamma_{gas}=5/3$. At the center of A2276, $T\simeq 3$ keV, and hence $c_{s}\simeq 895$ km/s, which implies that the radio bubble is effectively rising sub-sonically.

Since the cavity is located at about $r\simeq$21 -- 69 kpc from the cluster center, we obtain a buoyancy age of about $t_{b}\simeq 65\pm 46$ Myr 
by assuming a constant rising speed. Despite the large uncertainties involved, this age is on the same order as the radiative age of $t_{rad}\simeq 86 $ Myr estimated by Murgia et al. (2011) from the analysis of the total radio spectrum of the dying source and by assuming $B_{min}$=10 $\mu$G. We investigate this point in more detail in Fig.\,8, where we show the comparison of radiative and buoyancy ages as a function of the source's
 magnetic field. The trend of the radiative age (green line and shaded region) is derived from 

\begin{equation}
t_{rad}= 1590 \frac{B^{0.5}} {(B^{2}+B_{\rm IC}^2)[(1+z)\nu_{\rm break}]^{0.5}}  ~~~~\rm (Myr),
\label{trad}
\end{equation}

\noindent 
where the magnetic field is measured in $\mu$G, $B_{\rm IC}=3.25(1+z)^2$ is the equivalent inverse Compton field associated to the cosmic background radiation, and 
 $\nu_{\rm break}\simeq 0.23_{-0.2}^{+1.5}$ GHz is the spectral break measured in the total radio spectrum by Murgia et al. (2011).
The horizontal blue dashed line is the buoyancy age for the south lobe. The two ages match at 65 Myr for $B\simeq 12 ~\mu$G. This value for the 
source's magnetic field fits within the equipartition range $B_{min}$= 5.5 -- 22.7 $\mu$G for $k=1$ pictured by the vertical shaded region in red. 
This means that if the source is in the minimum energy condition, the pressure equilibrium with the external medium, $P_{nth}\sim10^{-10}$ dyne/cm$^2$, cannot be attained even supposing that $k\gg 1$. In fact, the relativistic particles energy density should be on the same order of the magnetic field energy density, i.e., $(1+k) u_{el}\simeq u_{B}\sim 5.7\times 10^{-12}$ dyne/cm$^2$, see Eq.\ref{equi}. Indeed, a possibility is that $k\sim 20$ and the source is not in the minimum energy condition because the
 energy density is dominated by non-radiating relativistic particles.
Alternatively, the source is in the minimum energy condition with $k=1$ and the pressure equilibrium is sustained by thermal material entrained in the radio lobes.

Whatever the reason, by assuming a similar energy output for the north cavity, the total power deposited by the AGN at the center of A2276 is estimated to be about $Q_{AGN}=2\times 4 P_{th}/t_{b}\sim 5\times 10^{44}$ erg/sec. The energy power from the AGN outburst is almost one order of magnitude higher than
the X-ray luminosity of A2276. Indeed, it is sufficient that a small fraction of this power is dissipated in the intra-cluster medium to reheat 
the cool core of the cluster.

In ZwCl 1829.3+691 a cavity is detected in the west lobe of the dying source WNB 1829+6911 at a distance of about 40 kpc from the core. We performed a 
similar calculation as for A2276 and found a  buoyancy age of about $t_{b}\sim 58$ Myr. According to Murgia et al. (2011) the 
radiative age is much longer $t_{rad}\sim 218$ Myr. However, a detailed comparison is more difficult in this case since the dying source has a restarting core 
whose spectrum overlaps that of the fossil  outburst. Nevertheless, the total power deposited by the AGN is estimated to be about $Q_{AGN}\sim$ 2.6 -- 10 $\times 10^{44}$ erg/sec. Thus, in this case the energy power from the AGN outburst is also far in excess of the X-ray luminosity of the cluster and thus sufficient to 
reheat the cool core.

\section{Conclusion}

We presented X-ray observations performed with the Chandra satellite of the three galaxy clusters Abell 2276, ZwCl\,1829.3+6912, and RX\,J1852.1+5711, which
 harbor at their center a dying radio source with an ultra-steep spectrum.

We analyzed the physical properties of the X-ray emitting gas that surrounds these elusive radio sources and concluded that the
physical state of the intra-cluster medium in the three galaxy clusters presents quite similar properties.

The large-scale X-ray emission is regular and spherical, suggesting a relaxed state for these systems. In fact, we
found that the three systems are also characterized by strong enhancements in the central abundance and declining temperature profiles toward the
 central region. For all these reasons, we classify RX J1852.1+5711, Abell 2276, and ZwCl 1829.3+6912 as cool-core galaxy clusters.

We found only marginal evidence for the presence of X-ray cavities associated to the fossil radio lobes 
in two of our galaxy clusters, namely A2276 and ZwCl 1829.3+691.
 
We compared the pressure of the intra-cluster medium with the non-thermal pressure of the radio lobes assuming that the radio sources are in the minimum energy condition.
 For all dying sources we found that this is on average about one to two orders of magnitude lower than that of the external gas, as found 
for many other radio sources at the center of galaxy groups and clusters.

We estimated the outburst age and energy output for the dying sources in Abell 2276 and ZwCl 1829.3+691, the two
clusters for which we have a putative cavity detection.

For WNB 1734+6407, the dying source at the center of A2276, the radiative age obtained from the analysis of the radio spectrum matches
 the buoyancy age estimated from the X-ray analysis if the source is in the minimum energy condition. Indeed, the pressure equilibrium with the external medium
is reached because the energy density of non-radiating particles (either relativistic or at thermal energies) is much higher than the magnetic field energy density.

We calculated the mechanical work done by the radio bubbles on the external gas and divided by the age of the radio source. The energy power from the AGN outburst 
is significantly higher than the X-ray luminosity in both clusters. Indeed, it is sufficient that a small fraction of this power is dissipated in the intra-cluster medium 
to reheat the cool cores.

\begin{acknowledgements}
We acknowledge the referee for helpful and constructive comments that improved the paper.
FG and MM are grateful for the hospitality of the Harvard-Smithsonian Center for Astrophysics, where part of this
research was done. Support was provided by Chandra grant GO9-0133X, NASA contract NAS8-39073, and the Smithsonian
Institution.
The National Radio Astronomy Observatory is operated by Associated Universities, Inc., under contract with the National Science Foundation.
This research made use of the NASA/IPAC Extragalactic Database (NED) which is operated by the Jet Propulsion Laboratory, California Institute of Technology, under contract  with the National Aeronautics and Space Administration.
The optical DSS2 red images were taken from: http://archive.eso.org/dss/dss. Funding for the SDSS and SDSS-II has been provided by the Alfred P. Sloan Foundation, the Participating Institutions, the National Science Foundation, the U.S. Department of Energy, the National Aeronautics and Space Administration, the Japanese Monbukagakusho, the Max Planck Society, and the Higher Education Funding Council for England. The SDSS Web Site is http://www.sdss.org/.
The SDSS is managed by the Astrophysical Research Consortium for the Participating Institutions. The Participating Institutions are the American Museum of Natural History, Astrophysical Institute Potsdam, University of Basel, University of Cambridge, Case Western Reserve University, University of Chicago, Drexel University, Fermilab, the Institute for Advanced Study, the Japan Participation Group, Johns Hopkins University, the Joint Institute for Nuclear Astrophysics, the Kavli Institute for Particle Astrophysics and Cosmology, the Korean Scientist Group, the Chinese Academy of Sciences (LAMOST), Los Alamos National Laboratory, the Max-Planck-Institute for Astronomy (MPIA), the Max-Planck-Institute for Astrophysics (MPA), New Mexico State University, Ohio State University, University of Pittsburgh, University of Portsmouth, Princeton University, the United States Naval Observatory, and the University of Washington. This research made use of Montage, funded by the National Aeronautics and Space Administration's Earth Science Technology Office, Computation Technologies Project, under Cooperative Agreement Number NCC5-626 between NASA and the California Institute of Technology. Montage is maintained by the NASA/IPAC Infrared Science Archive.
\end{acknowledgements}

\appendix
\section{Minimum energy calculation for an isotropic electron population with a power law spectrum in a disordered magnetic field}

The total non-thermal energy density of the radio lobes is written as the sum of the ions, electrons, and magnetic field contributions:

\begin{equation}
u_{tot}=u_{ions}+u_{el}+u_{B}.
\end{equation}
The energy density in non-radiating particles is set simply proportional to that of electrons

\begin{equation}
u_{tot}=(1+k) u_{el}+u_{B}.
\end{equation}

The electron energy spectrum (number density of electrons with a Lorentz factor between $\gamma$ and $\gamma+d\gamma$) is taken to be a power law 

\begin{equation}
N(\gamma,\theta)=K_0 \gamma^{-p}(\sin\theta)/2,
\end{equation}
where the electron's Lorentz factor  $\gamma$ is in the range from $\gamma_{min}$ to $\gamma_{max}$  and we assumed an isotropic distribution of the pith angle $\theta$ between the local direction of the magnetic field and the electron velocity. Thus, the electron energy density is

\begin{equation}
u_{el}=m_{e}c^2 K_0/2 \int_{\gamma_{min}}^{\gamma_{max}}\int_{0}^{\pi}\gamma^{-p+1}\sin\theta d\gamma d\theta.
\label{uel}
\end{equation}

The radio emissivity as a function of the pitch angle in a uniform magnetic field of strength $B$ is given by
\begin{equation} 
j_{syn}(\nu,\theta)=\int_{\gamma_{min}}^{\gamma_{max}}C_f B\sin\theta\,N(\gamma,\theta)F_{syn}(\nu/\nu_{c})d\gamma ~~~{\rm \left(\frac{erg}{cm^{3}s\,Hz}\right)},
\label{jradiotheta}
\end{equation}
where $F_{syn}(\nu/\nu_{c})$ is the synchrotron kernel (see e.g.  Blumenthal \& Gould 1970; Rybicki \& Lightman 1979), while
  
\begin{equation} 
\nu_{c}=C_{\nu} B \sin\theta\,\gamma^2    ~~~\rm (Hz).
\label{nusyn}
\end{equation}
The constants $C_f=2.3444\times 10^{-22}$ (erg Gauss$^{-1}$) and $C_{\nu}=4.1989\times 10^{6}$ (Gauss$^{-1}$s$^{-1}$) depend only on fundamental
physical constants.

We now suppose that the magnetic field is completely tangled in an infinitesimally
 small scale compared to the source's size. With this assumption, the synchrotron
luminosity, after averaging Eq.\ref{jradiotheta} over all possible magnetic field directions with respect to the line-of-sight (LOS), is

\begin{equation} 
L_{\nu}=V \int_{0}^{\pi} j_{syn}(\nu,\theta^{\prime}) \frac{\sin\theta^{\prime}}{2} d\theta^{\prime},
\label{jradio}
\end{equation}
where $V$ is the source's volume, see also the discussion in Murgia et al. (2010). In the power law regime, $\nu_{min}\ll \nu \ll \nu_{max}$, we obtain the
well-known formula for the synchrotron monochromatic power

\begin{equation} 
L_{\nu}=\langle C_{\alpha} \rangle_{LOS} V K_{0} B^{\alpha+1} \nu^{-\alpha},
\label{jradioLOS}
\end{equation}
where $\alpha=(p-1)/2$ while

\begin{equation} 
\langle C_{\alpha} \rangle_{LOS}=\frac{C_{\nu}^{\alpha}C_{f}}{8} \int_{0}^{\pi} (\sin\theta)^{3+\alpha}d\theta \int_{0}^{\infty} x^{\alpha-1}F(x)dx.
\label{calpha}
\end{equation}
In the literature an idealized situation is often assumed in which the source's  magnetic field lies in the plane of the sky. 
In this case, due to the high beaming of the synchrotron radiation, 
only electrons with pith angle $\theta=90\degr$  can be observed and the constant in Eq.\ref{jradioLOS} has to be replaced by

\begin{equation} 
(C_{\alpha})_{90\degr}=\frac{C_{\nu}^{\alpha}C_{f}}{4} \int_{0}^{\infty} x^{\alpha-1}F(x)dx.
\label{calpha}
\end{equation}
The synchrotron radiation is then amplified by 60$-$70\% with respect to the more realistic situation in which 
the field direction is disordered in the source. 

For $\alpha$ in the range between 0.5 and 1.0, Eq.\,A.9 can be approximated by the polynomial expansion:

\begin{equation} 
\langle C_{\alpha} \rangle_{LOS}\approx \frac{C_{\nu}^{\alpha}C_{f}}{8} (2.277-2.616\alpha+1.317\alpha^2).
\label{calpha}
\end{equation}

Substituting Eq.\ref{jradioLOS} in Eq.\ref{uel}, we found that the source's non-thermal energy is
 minimum for

\begin{equation} 
B_{min}=\left[\frac{4\pi m_{e}c^2 (1+\alpha) (1+k) \nu^{\alpha}}{\langle C_{\alpha} \rangle_{LOS} }\left(\frac{\gamma_{max}^{1-2\alpha}-\gamma_{min}^{1-2\alpha}}{1-2\alpha}\right) \frac{L_{\nu}}{V}\right] ^{1/(3+\alpha)}. 
\label{bmin}
\end{equation}
For this value of the magnetic field strength the energy is nearly equipartioned between particles and field:
 
\begin{equation} 
(1+k) u_{el} =\left( \frac{2}{1+\alpha} \right) u_{B},
\label{equi}
\end{equation}

so that the total minimum energy density is

\begin{equation} 
u_{min}=\left( \frac{3+\alpha}{1+\alpha} \right) \frac{B_{min}^2}{8\pi}.
\end{equation}

\end{document}